\documentstyle[twoside,fleqn,espcrc2]{article}

\newcommand{\AmS}{{\protect\the\textfont2
  A\kern-.1667em\lower.5ex\hbox{M}\kern-.125emS}}
\title{Dynamical Supersymmetry Breaking}
\author{Ann E. Nelson\thanks{Research supported in part by the DOE under grant 
                \#DE-FG03-96ER40956. Preprint no. UW-PT/97-20.}
\address{Department of Physics,
        University of Washington, \\ 
        P.O. Box 351560, Seattle, WA 98195-1560 USA}} 
\begin{document}
\begin{abstract} I review the motivation for dynamical supersymmetry
  breaking, the various mechanisms which have been discovered, and the
  prospects for model building.
\end{abstract}
\maketitle

\section{Motivation: The Hierarchy Problem}

The ratio of the weak scale to the Planck scale, $m_W/m_{Pl}\sim
10^{-17},$
suggests that it should have a dynamical origin. Had this ratio been,
{\it e.g.},
$\sim 10^{-2}$ we would have suspected that the weak scale
is a perturbative effect. In fact this ratio is of the right  size for
a nonperturbative effect in field theory. 
Had it been {\it extremely} small, {\it e.g.} 
$\sim10^{-10^{10}}$ or so, we would have to look for some really
exotic, unprecedented explanation.

Whereas currently the hierarchy problem is the puzzle of why the weak
scale is so small, I hope that within 10 years or so we will learn 
that the weak scale is tied to a low supersymmetry breaking 
(SSB) scale, and the
hierarchy problem will be reformulated as  the question of why the SSB
scale is so low. The likely answer to this question is that SSB is a
nonperturbative effect proportional to $e^{-4\pi/b\alpha}$, where $b$
is an integer
of order one and $\alpha$ is some coupling which is weak at
$m_{Pl}$.

\section{Recent Advances in Dynamical Supersymmetry Breaking}

Until c. 1994, the most popular approach to SSB was to assume that it
came from some ``hidden'' sector and treat its consequences
phenomenologically, ignoring the underlying mechanism. 
One reason for this was that explicit models of
dynamical SSB  (DSB) had several unattractive features.

Successful explicit renormalizable models 
    either required very large gauge groups or a contrived looking
  modular approach, with three separate sectors for DSB, the MSSM, and
  the
  ``messengers'' which share gauge interactions with both the DSB and
  MSSM sectors.

Implementing DSB in a hidden sector which only couples to the
  MSSM via Planck scale effects seemed to imply the existence of very
  light gauginos. Gaugino masses in this scenario arise from the terms in
  the 
  gauge kinetic function
  \begin{equation}
  \int d^2\theta f(\Phi){\cal W}_\alpha{\cal W}^\alpha \ ,
  \end{equation}
  where $f(\Phi)$ is a gauge invariant polynomial of chiral superfields.
  However known successful DSB models were completely chiral under the
  gauge interactions, and so  $f(\Phi)$ had to be of at least cubic
  order, implying that the visible gaugino masses  were proportional to
  at least three powers of $1/m_{Pl}$ and were much smaller than the
  squark and slepton masses.  

Fortunately, the work of Seiberg, Seiberg-Witten and co. 
\cite{seiberg,SW} has
dramatically improved our understanding of nonperturbative effects in
supersymmetric theories and lead to many new mechanisms for and models
of DSB \cite{skiba}.
Modern DSB models are known which
\begin{enumerate}
\item contain gauge singlets, hence are quite workable as hidden
  sector models with comparable gaugino and scalar masses \cite{singlets,qms}.
\item are not chiral under the gauge group \cite{qms}. 
This is possible in theories where adding mass terms to all
superfields allows  supersymmetric vacua to move in from 
infinity in field space.
\item have large unbroken global symmetries, enabling direct embedding of
  the standard model gauge interactions into the DSB sector
  \cite{nomess}.
\item have classically flat directions lifted by quantum
  effects \cite{singlets,qms,nomess,flat}. As I will discuss later, this is
  a necessary condition for the quarks and leptons to participate
  directly in the supersymmetry dynamics.
\item generate multiple scales \cite{singlets,qms,nomess,flat,dnns}.  
This has a variety of model building  applications. 
\item do not have any $U(1)_R$ invariance \cite{noR}.
\end{enumerate}
No models with any of these features were known prior to 1994. In
addition, several new mechanisms have been discovered. Furthermore,
examples are known in which 
the various  mechanisms occur in a dual description of the
theory \cite{IT}.

\section{Review of DSB Mechanisms}
\subsection{Strong Coupling}

In general, a theory with no classically flat directions and a broken
global continuous symmetry will break supersymmetry, according to the
following argument: Unbroken
supersymmetry requires that a Goldstone boson reside in a
supermultiplet whose scalar component corresponds to a noncompact
flat direction of the potential. The first
examples of DSB fell into this category \cite{sufive}. 

Strongly coupled DSB theories have no small parameters.
For instance consider the gauge
group is $SU(5)$, with chiral superfields $\bar 5+10$. This theory
admits no gauge invariant superpotential whatsoever, and has no
classically flat directions. The anomaly free global symmetry is
$U(1)_R\times U(1)_{B-L}$. Because of the difficulty of matching the
global anomalies, it was assumed that at least part of the global
symmetry is spontaneously broken, and so supersymmetry also must
break \cite{sufive}.

Recently more evidence for DSB has been gathered for the SU(5) theory
and also for SO(10)
with a single spinor, using the following technique:
Add vector-like matter to the theory with a mass 
$M$ \cite{addvector}. In the
large $M$ limit, this addition should not affect the dynamics. Now
analyticity arguments \cite{seiberg,witten} can be used to show that
whether or not supersymmetry is broken does not depend on 
the size of $M$, for $M$ non-zero. In the limit where $M$ is small,
supersymmetry breaking can be demonstrated in a
weakly coupled dual description.

\subsection{Dynamically Generated Superpotential}
 A good example of supersymmetry breaking which can be studied at weak
 coupling \cite{calculable} is based on the gauge group 
$SU(3)\times SU(2)$ 
\cite{ADS},
 with matter content 
\begin{equation}
q\sim(3,2),\ \ \bar u\sim(\bar 3,1),\ \ \bar d\sim(\bar 3,1),\ \ \ell\sim(1,2)
\end{equation}
and tree level superpotential
\begin{equation}
W_{tree}=\lambda q\bar d\ell\ .
\end{equation}

This theory has no classically flat directions, 
and an exact $U(1)_R$ symmetry.

In the limit $\Lambda_3\gg\Lambda_2$,  the supersymmetry breaking is
dominated by $SU(3)$ instantons, which generate an effective  superpotential
\begin{equation}
W_{dyn}=\frac{\Lambda_3^7}{q^2\bar u\bar d}\ .
\end{equation}

This superpotential drives the scalar fields away from the origin of field
space and
leads to $U(1)_R$ and supersymmetry breaking. For $\lambda\ll 1$, the
minimum occurs at a point where where all the gauge groups are in a
weakly coupled Higgs phase and so quantitative calculations are possible.

\subsection{Quantum Modified Constraint}

The phenomenom of quantum modification of the moduli space
\cite{seiberg} can also lead to DSB \cite{qms}. A good example of this
occurs in the same $SU(3)\times SU(2)$ model in the limit 
$\Lambda_2\gg\Lambda_3$. The $SU(2)$ theory with 4 doublets has a
quantum moduli space on which the $U(1)_R$ symmetry and thus 
supersymmetry are broken.

\subsection{Confinement}

Intriligator, Seiberg and Shenker \cite{ISS} demonstrated that
confinement can lead to SSB in the following simple model.

Consider $SU(2)$ with a single chiral superfield $\psi$ in an $I=\frac32$
representation, and superpotential. 

\begin{equation}
{\cal W}=\frac\lambda{m_{Pl}}\psi^4\ .
\end{equation}

Classically this theory has a unique supersymmetric minimum at the
origin of field space. It is not possible to dynamically generate a
superpotential.

The anomalies of the theory are matched by a single gauge invariant 
chiral superfield
$\phi=\psi^4$. Provided this $SU(2)$ theory confines, the K\"ahler
metric is singular in terms of the $\psi$ variable but smooth  and
positive definite in terms of $\phi$. Thus

\begin{equation}
V=K_{\phi\phi^*}^{-1}\left|\frac{\partial W}{\partial \phi}\right|^2 >0
\end{equation}
and supersymmetry is broken.

\subsection{Product Group Mechanisms}

Supersymmetry can also be broken by Rube-Goldberg-like mechanisms
involving nonperturbative dynamics in more than one gauge group
\cite{RubeGoldberg}. 

Many such models have been obtained by ``reducing'' known DSB models,
{\it i.e.} throwing away some of the gauge generators \cite{dnns}.

For example the supersymmetry breaking $SU(2N+1)$ theory with an 
antisymmetric tensor chiral 
superfield and
$(2N-3)$ anti-fundamentals \cite{ADS}, can be reduced to
$SU(m)\times SU(2N-m+1)\times U(1)$,  with the matter content
given by the decomposition of the $SU(2N+1)$ superfields under this
subgroup. Leigh, Randall and Rattazzi (LRR) \cite{LRR} have given a clever
argument to show why this particular decomposition will lead to a new
DSB model. They consider adding a superfield $A$  in the adjoint
representation to the parent $SU(2N+1)$ model, and a superpotential
containing the terms
\begin{equation}
\frac12 m A^2+\frac13\lambda A^3\ .
\end{equation}

The resulting theory has minima under which the effective theories are
the ``daughter'' models with gauge group 
$SU(m)\times SU(2N-m+1)\times U(1)$. By considering a dual of the
parent theory, LRR are able to show that it has no supersymmetric
minima, and therefore neither do any of the daughter theories.

In many of the daughter theories, complicated nonperturbative dynamics
in two gauge factors are needed to break supersymmetry.
For example, in the $SU(n)\times SU(5)\times
U(1)$, theory, the $SU(5)$ factor, would, in
isolation, be in a conformal phase (for $n>5$), and the $SU(n)$ in a
free magnetic phase, both phases allowing a supersymmetric minimum at the 
origin. However, the simultaneous presence of a tree level
superpotential and the dynamics of both groups drives the theory into
a SSB phase. One can see DSB by considering a dual of the $SU(5)$
under which the tree level superpotential maps onto mass terms for
some of the $SU(n)$ flavors. These mass terms  drive the $SU(n)$ 
theory into a phase in which a superpotential is dynamically generated,
excluding  the
origin, and spontaneously breaking a $U(1)_R$ symmetry and
supersymmetry.

\section{New Possibilities for Model Building}

The first attempts at building realistic models with low energy DSB
required very large gauge groups ({\it e.g.} $SU(15)$) in
the DSB sector, which meant that the ordinary $SU(3)\times SU(2)\times
U(1)$ gauge couplings had Landau poles at relatively low 
energies \cite{ADS}.
This problem was solved by introducing a ``messenger sector''
communicating with the DSB sector via a new, ``messenger'' gauge
group, and with the MSSM sectors via the usual $SU(3)\times SU(2)\times
U(1)$ interactions \cite{dnns,dn}. However the messenger sector and
gauge interactions appear  unnattractively {\it ad hoc}. It would be
more elegant  if the messenger particles and $SU(3)\times SU(2)\times
U(1)$ interactions were an organic part of
the DSB sector. Recent understanding of nonperturbative effects in
supersymmetric gague theories have enabled 
much model building progress
on simplifying or eliminating the need for a separate messenger sector
\cite{qms,nomess}.

An  intriguing possibility is that some of the quarks and leptons
could participate in the DSB dynamics directly \cite{CKN,luty}.
The corresponding superpartners would gain supersymmetry breaking 
masses which are larger
than the gaugino masses by a factor of 
$\sim{\cal O}\left(16\pi/g^2\right)$,
where $g$ is the relevant $SU(3)\times SU(2)\times
U(1)$ gauge coupling \cite{CKN,fourpi}. Thus, assuming that the
gluino mass should be $\sim{\cal O}(100$~GeV) or more, some squark and
slepton masses would be in the $\sim{\cal O}(10$~TeV) range
\cite{CKN}. Such  a heavy mass for the top and left handed bottom
squarks, or for the up-type Higgs,  would pose a severe problem for 
natural electroweak symmetry
breaking, so we conclude that the Higgs and its close relatives must
remain isolated from the DSB sector. However, such heavy masses
for the first and second generations
of squarks and sleptons do not necessarily upset natural electroweak
symmetry breaking, since these are quite weakly coupled to the 
Higgs \cite{CKN,dg,cousins}. 

There are, however, several difficulties with this scenario. 
The most severe
problem is that it is still not known whether any 
such viable model with a stable supersymmetry breaking
minimum exists. The many classically flat directions of the 
MSSM would have to 
be lifted by DSB dynamics. Fortunately, examples are now known of
models with classically flat directions which are all 
lifted by quantum mechanical
effects after supersymmetry is broken. DSB examples are also known
with light composite fermions, although I know of no such
examples where those fermion quantum numbers could be consistent with 
the quark or lepton $SU(3)\times SU(2)\times
U(1)$ charges. However I suspect that such DSB models do exist and will
be found in the next few years.

A second potential problem is that integrating out the heavy squarks and
sleptons 
could generate
an excessively large Fayet Iliopoulos term for hypercharge
\cite{dg}. This problem could be solved if $SU(3)\times SU(2)\times
U(1)$ is embedded into an approximate global $SU(5)$ symmetry of the DSB sector
\cite{CKN}.

A final difficulty is that the squarks and sleptons now need not be
approximately degenerate, since they do not learn about supersymmetry
breaking exclusively via their $SU(3)\times SU(2)\times
U(1)$ interactions. Thus the spectre of flavor changing
neutral currents and lepton flavor violation is raised. 
However these flavor
violating effects may be acceptably small, provided that those members
of the first two generations with the same  $SU(3)\times SU(2)\times
U(1)$ charges are both either strongly involved in DSB and so very
heavy or coupled to DSB mainly via $SU(3)\times SU(2)\times
U(1)$ gauge interactions and so nearly degenerate. 

Given these considerations, Fayet Iliopoulos terms and flavor
violation may be adequately suppressed in several different
ways \cite{hybrid}. For instance the scalar members of the 
first two generations of
10's (under a global $SU(5)$) could be heavy, or the $\bar 5$
scalars could be heavy,
or both. One could consider, {\it e.g.} the unusual possibility
of relatively light and degenerate left handed selectron and smuon but
very heavy right handed sleptons for the first two generations, which
could have interesting implications for LEP II \cite{hybrid}.

Such mingling of messenger, MSSM and DSB sectors could have
significant advantages as well, conceivably explaining the fermion
mass hierarchy, the size of the $\mu$ parameter and the absence of baryon
and lepton number violation \cite{CKN}. We believe that new 
nonperturbative gauge interactions are necessary for DSB, and it would
be beautiful if such gauge symmetries could solve these other problems of
supersymmetric theories. 

\section{Conclusions}

Clearly we have made tremendous progress in the last three years 
in understanding dynamical supersymmetry breaking. However at least
two intriguing questions remain unanswered:

\begin{itemize}
\item Is there a simple algorithm for determining which theories
  exhibit DSB? If so, what is it?
\item Could some of the quark and lepton superfields or ordinary gauge
  interactions be an integral part of the DSB dynamics?
 \end{itemize}

\end{document}